\documentclass[12pt]{iopart}
\usepackage{iopams}
\usepackage{graphicx}

\begin{document}

\title[Brownian particle in inhomogeneous media and driven by colored noise]{Power
law statistics in the velocity fluctuations of Brownian particle in inhomogeneous
media and driven by colored noise}

\author{R Kazakevi\v{c}ius and J Ruseckas}

\address{Institute of Theoretical Physics and Astronomy, Vilnius University,
A. Go\v{s}tauto 12, LT-01108 Vilnius, Lithuania}
\ead{rytis.kazakevicius@tfai.vu.lt}

\begin{abstract}
In this paper we consider the motion of a Brownian particle
in an inhomogeneous environment such that the motion can be described
by the equation yielding $1/f$ spectrum in a broad range of frequencies.
The inhomogeneous environment can be a result, for example, of a linear
potential affecting the Brownian particle together with the medium
where steady state heat transfer is present due to the difference
of temperatures at the ends of the medium. The correlation of collisions
between the Brownian particle and the surrounding molecules can lead
to the situation where the finite correlation time becomes important,
thus we have investigated the effect of colored noise in our model.
Existence of colored noise leads to the additional restriction of
the diffusion and exponential cut-off of the distribution of particle
positions. Narrower power law part in the distribution of the particle
positions results in the narrower range of frequencies where the spectrum
has power law behavior.

\end{abstract}


\submitto{{\it J. Stat. Mech.}}

\maketitle

\section{Introduction}

One of the characteristics of the noise is the power spectral density
(PSD). The white noise has a frequency-independent PSD, whereas the
PSD of a colored noise depends on the frequency, the characteristic
behavior of the PSD is referred to as a ``color'' of the noise. There
are various applications where the noise in the physical system under
investigation has a non-trivial spatio-temporal
structure and where it is not realistic to model it is as a white
noise process. For example, for a Brownian particle the driving noise
is actually colored, i.e.\ it has a characteristic non-zero correlation
time $\tau$ on a short time-scale of the order of tens of nanoseconds
\cite{Franosch2011}. Noise color arises due entrainment of fluid
around diffusing particle. The particle accelerates the entrained
fluid and this acceleration depends on the past motion of the particle
and introduce an inertial memory effect \cite{Landau1987}.

Theoretical models suggests that variety of systems affected by a
colored noise instead of a white noise exhibit new interesting properties.
For example: The correlations in colored noise are found to be able
to enhance or suppress the growth rate of amplification above or below
a critical detuning in the collective scattering of light from a laser
with a colored noise in ultracold and collisionless atomic gas \cite{Zhou2009}.
Investigation of the colored-noise effect on nonequilibrium phase
transitions shows reentrant transitions from ordered into the disordered
phase as the correlation time and the coupling strength increase \cite{Kim1998}.
The color and coupling induced disorders are pure colored-noise effects
because of the absence of the white-noise limit.
Some of the population growth models subjected to a white environmental
noise changes the population-size dependence of the mean time to extinction
(MTE) from an exponential to a power-law with a large exponent \cite{Lande1993}.
The introduction of the colored Gaussian noise changes this exponent,
reducing it at a fixed noise magnitude. For a long correlation time
of the environmental noise the MTE becomes independent of the population
size for a strong enough noise \cite{Kamenev2008}.

Investigation of the effects caused by the presence of a colored noise
in physical systems has some practical implications. Study of colored-noise-induced
synchronization in chaotic systems indicates that the critical amplitude
required for synchronization is generally smaller for the white noise
as compared with the colored noise \cite{Wang2009}. A practical implication
is that, in situations where synchronization is undesirable, a simple
control strategy is to place filters in the system so as to make the
noise source as colored as possible. In the systems exhibiting the
phenomenon of stochastic resonance an exponentially correlated noise
(``red'' noise) leads to a reduction of signal amplification and the
peak of stochastic resonance moves to a larger noise intensity when
the correlation time increases \cite{Gammaitoni1998}. ``Pink noise''
or $1/f$ noise, as opposed to white noise also leads to a reduction
of signal amplification, but resonance peak arises for lower noise
intensity, if special conditions are satisfied \cite{Nozaki1999}.
This is important for understanding weak signal transmission trough
noisy environments.

Signals having the PSD at low frequencies $f$ of the form $S(f)\sim1/f^{\beta}$
with $\beta$ close to $1$ are commonly referred to as ``pink noise''
or ``$1/f$ noise''. Such signals are often found in physics and in
many other fields \cite{Scholarpedia2007,Weissman1988,Barabasi1999,Gisiger2001,Wagenmakers2004,Szabo2007,Castellano2009}.
Since the discovery of $1/f$ noise numerous models and theories have been
proposed, for a recent review see \cite{Balandin2013}. Mostly $1/f$ noise is
considered as a Gaussian process \cite{Kogan2008,Li2012}, but sometimes the
$1/f$ fluctuations are non-Gaussian \cite{Orlyanchik2008,Melkonyan2010}. The
Brownian motion as a source of $1/f$ noise was first proposed in the seminal
paper by Marinari \textit{et al.} \cite{Marinari1983}, where it was suggested that
$1/f$ noise can result from a random walk in a random
environment. Starting from the model of $1/f$ noise as a
Brownian motion of inter-pulse durations
\cite{Kaulakys1998,Kaulakys1999,Kaulakys2005}, nonlinear SDEs generating signals
with $1/f$ spectrum have been derived in~\cite{Kaulakys2004,Kaulakys2006}.
A special case of this nonlinear SDE has been obtained using Kirman's agent
model \cite{Ruseckas2011epl}. Such nonlinear SDEs have been used to describe
signals in socio-economical systems \cite{Gontis2010,Mathiesen2013}. In this
paper we consider the motion of a Brownian particle in an inhomogeneous
environment and described by the nonlinear SDE yielding $1/f$ spectrum.

The influence of the colored noise on the dynamics of a Brownian particle
immersed in a fluid where a temperature gradient is present can lead
to interesting phenomena. The particle can exhibit a directed motion
in response to the temperature gradient. Furthermore, study of stationary
particle distribution shows that particles can accumulate towards the
colder (positive thermophoresis) or the hotter (negative thermophoresis)
regions depending on their physical parameters and, in particular,
on the dependence of their mobility on the temperature \cite{Hottovy2012}.
The velocity of this motion can vary both in magnitude and sign, as
observed in experiments \cite{Piazza2008a}. However, in this case,
no external force is actually acting on the particles \cite{Piazza2008}.
Theoretical models suggest \cite{Hottovy2012} the presence of a colored
noise, as opposed to a white noise, is crucial for the emergence of
such thermophoretic effects. Analysis of the steady-state dynamics
of an overdamped classical particle in asymmetric multidimensional
potential driven by the noise with an arbitrary correlation function has
shown that the correlated noise breaks detailed balance, thereby exploiting
the spatial asymmetries in potential to produce local drifts and rotations
\cite{Ghosh2000}. These interesting findings motivated us to investigate
the motion of a Brownian particle in an inhomogeneous environment
and subjected to a colored noise, as opposed to a white noise.

The paper is organized as follows: In \sref{sec:nonlinear-SDE}
we review nonlinear SDEs driven by a white noise and yielding power-law
steady state probability density function (PDF) of the generated signal.
We estimate when the signal generated by such an SDE has $1/f$ PSD
in a wide region of frequencies. In \sref{sec:Deriv_mod}
we show that nonlinear SDEs generating power-law distributed precesses
with $1/f^{\beta}$ spectrum can result from diffusive particle motion
in inhomogeneous medium. In \sref{sec:colored} we present
methods that we use to study the influence of colored noise and in
\sref{sec:col_on_model} we investigate the effect of colored
noise in our model: numerically solve obtained equations and compare
the PDF and PSD of the signal with analytical estimations. \Sref{sec:concl}
summarizes our findings.

\section{Nonlinear stochastic differential equations generating signals with
$1/f$ spectrum}

\label{sec:nonlinear-SDE}The nonlinear SDEs generating power-law
distributed precesses with $1/f^{\beta}$ noise have been derived
in papers \cite{Kaulakys2004,Kaulakys2006,Kaulakys2009} starting
from the point process model \cite{Kaulakys1998,Kaulakys1999,Kaulakys2005}.
The general expression for the proposed class of It\^o SDEs is
\begin{equation}
\rmd x=\sigma^{2}\left(\eta-\frac{1}{2}\lambda\right)x^{2\eta-1}\rmd t
+\sigma x^{\eta}\rmd W_{t}\,.\label{eq:sde-ito}
\end{equation}
Here $x$ is the signal, $\eta\neq1$ is the exponent of the power-law
multiplicative noise, $\lambda$ defines the exponent of the power-law
steady-state PDF of the signal, $W_{t}$ is a standard Wiener process
(the Brownian motion) and $\sigma$ is a scaling constant determining
the intensity of the noise. The Fokker-Planck equation corresponding
to SDE \eref{eq:sde-ito} gives the power-law probability density
function (PDF) of the signal intensity $P_{0}(x)\sim x^{-\lambda}$
with the exponent $\lambda$. The non-linear SDE \eref{eq:sde-ito}
has the simplest form of the multiplicative noise term,
$\sigma x^{\eta}\rmd W_{t}$.
In papers \cite{Ruseckas2011epl,Kononovicius2012} the nonlinear
SDE of type \eref{eq:sde-ito} has been obtained starting from a
simple agent-based model describing the herding behavior.

It\^o SDEs are typically used in economics \cite{Jeanblanc2009}
and biology \cite{Turelli1977}. On the other hand, Stratonovich SDEs
are more suitable for real systems with correlated, non-white noise,
for example, for noise-driven electrical circuits \cite{Smythe1983}.
The Stratonovich SDE corresponding to It\^o SDE \eref{eq:sde-ito}
is \cite{Gardiner2004} 
\begin{equation}
\rmd x=\frac{1}{2}\sigma^{2}(\eta-\lambda)x^{2\eta-1}\rmd t
+\sigma x^{\eta}\circ \rmd W_{t}\,.\label{eq:sde-stratonovich}
\end{equation}
Note, that the choice of Stratonovich or It\^o convention depends
not only on the correlation time of the noise but also on the delay
in the feedback \cite{Pesce2013}.

For $\lambda>1$ the distribution $P_{0}(x)$ diverges as $x\rightarrow0$,
therefore the diffusion of the stochastic variable $x$ should be
restricted at least from the side of small values, or equation \eref{eq:sde-ito}
should be modified. The simplest choice of the restriction is the
reflective boundary conditions at $x=x_{\mathrm{min}}$ and $x=x_{\mathrm{max}}$.
Another choice would be modification of equation \eref{eq:sde-ito} to
get rapidly decreasing steady state PDF when the stochastic variable
$x$ acquires values outside of the interval $[x_{\mathrm{min}},x_{\mathrm{max}}]$.
For example, the steady state PDF
\begin{equation}
P_{0}(x)\sim\frac{1}{x^{\lambda}}\exp\left(-\frac{x_{\mathrm{min}}^{m}}{x^{m}}
-\frac{x^{m}}{x_{\mathrm{max}}^{m}}\right)
\end{equation}
with $m>0$ has a power-law form when $x_{\mathrm{min}}\ll x\ll x_{\mathrm{max}}$
and exponential cut-offs when $x$ is outside of the interval
$[x_{\mathrm{min}},x_{\mathrm{max}}]$.
Such exponentially restricted diffusion is generated by the SDE
\begin{equation}
\rmd x=\sigma^{2}\left[\eta-\frac{1}{2}\lambda
+\frac{m}{2}\left(\frac{x_{\mathrm{min}}^{m}}{x^{m}}
-\frac{x^{m}}{x_{\mathrm{max}}^{m}}\right)\right]x^{2\eta-1}\rmd t
+\sigma x^{\eta}\rmd W_{t}\label{eq:4}
\end{equation}
obtained from equation \eref{eq:sde-ito} by introducing the additional
terms in the drift.

The PSD of the signals generated by the SDE \eref{eq:sde-ito} can
be estimated using the (approximate) scaling properties of the signals,
as it is done in \cite{Ruseckas2014}. Since the Wiener process
has the scaling property $dW_{at}=a^{1/2}dW_{t}$, changing the variable
$x$ in equation \eref{eq:sde-ito} to the scaled variable $x_{s}=ax$
or introducing the scaled time $t_{s}=a^{2(\eta-1)}t$ one gets the
same resulting equation. Thus change of the scale of the variable
$x$ and change of time scale are equivalent, leading to the following
scaling property of the transition probability (the conditional probability
that at time $t$ the signal has value $x'$ with the condition that
at time $t=0$ the signal had the value $x$):
\begin{equation}
aP(ax',t|ax,0)=P(x',a^{\mu}t|x,0)\,,\label{eq:trans-scaling}
\end{equation}
with the exponent $\mu$ being $\mu=2(\eta-1)$. As has been shown
in \cite{Ruseckas2014}, the power-law steady state PDF $P_{0}(x)\sim x^{-\lambda}$
and the scaling property of the transition probability \eref{eq:trans-scaling}
lead to the power-law behavior of the PSD
\begin{equation}
S(f)\sim\frac{1}{f^{\beta}}\,,\qquad\beta=1+\frac{\lambda-3}{2(\eta-1)}\label{eq:beta}
\end{equation}
in a wide range of frequencies.

The presence of the restrictions of diffusion at $x=x_{\mathrm{min}}$
and $x=x_{\mathrm{max}}$ makes the scaling \eref{eq:trans-scaling}
not exact and this limits the power-law part of the PSD to a finite
range of frequencies $f_{\mathrm{min}}\ll f\ll f_{\mathrm{max}}$.
The frequency range where the PSD has $1/f^{\beta}$ behavior is estimated
in \cite{Ruseckas2014} as 
\begin{eqnarray}
\sigma^{2}x_{\mathrm{min}}^{2(\eta-1)} \ll 2\pi f
\ll\sigma^{2}x_{\mathrm{max}}^{2(\eta-1)}\,,\qquad\eta>1\,,\label{eq:approx-range}\\
\sigma^{2}x_{\mathrm{max}}^{-2(1-\eta)} \ll 2\pi f
\ll\sigma^{2}x_{\mathrm{min}}^{-2(1-\eta)}\,,\qquad\eta<1\,.\nonumber 
\end{eqnarray}
This equation shows that the frequency range grows with increasing
of the exponent $\eta$, the frequency range becomes zero when $\eta=1$.
By increasing the ratio $x_{\mathrm{max}}/x_{\mathrm{min}}$ one can
get arbitrarily wide range of the frequencies where the PSD has $1/f^{\beta}$
behavior. Note, that pure $1/f^{\beta}$ PSD is physically impossible
because the total power would be infinite. Therefore, we consider
signals with PSD having $1/f^{\beta}$ behavior only in some wide
intermediate region of frequencies, $f_{\mathrm{min}}\ll f\ll f_{\mathrm{max}}$,
whereas for small frequencies $f\ll f_{\mathrm{min}}$ the PSD is
bounded.

\begin{figure}
\includegraphics[width=0.45\textwidth]{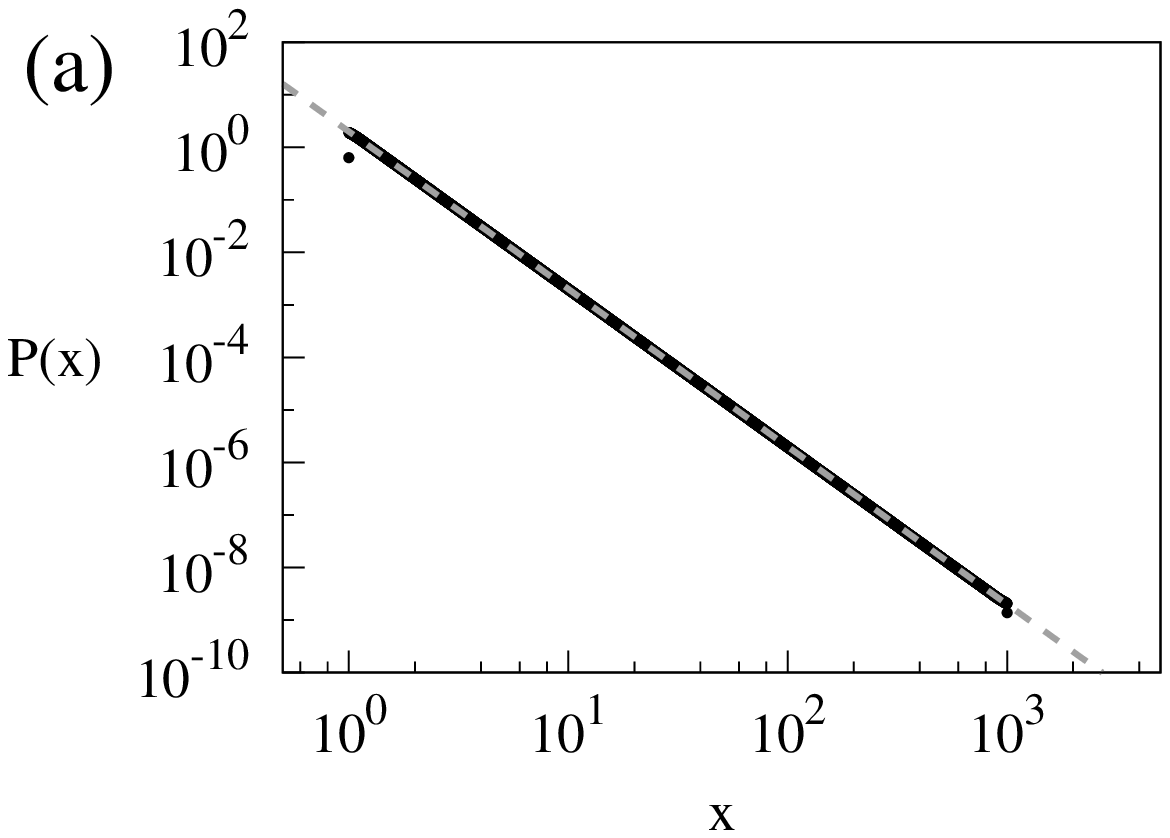}\includegraphics[width=0.45\textwidth]{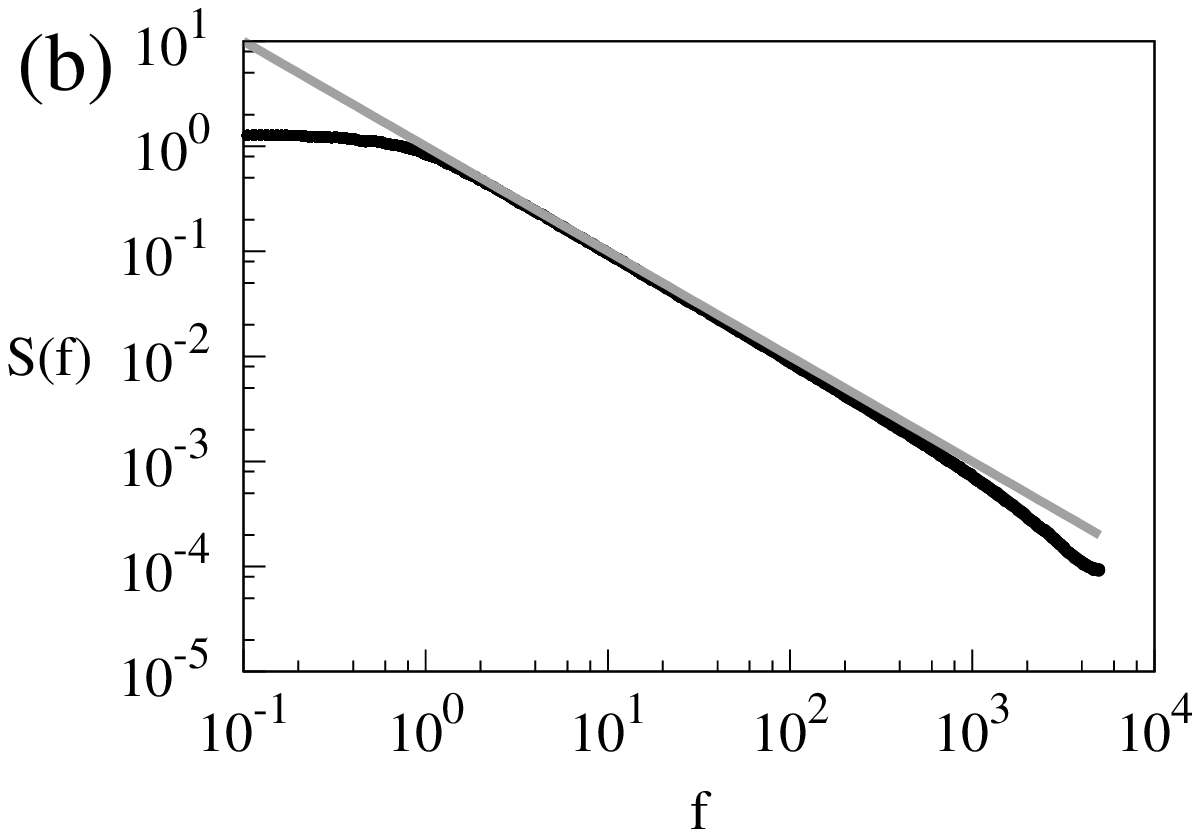}
\caption{(a) The steady-state PDF of the signal generated by equation \eref{eq:sde-ito}
with reflective boundaries at $x_{\mathrm{min}}$ and $x_{\mathrm{max}}$.
The dashed line shows the power-law with the exponent $-3$. (b) The
PSD of such a signal. The gray line shows the slope $f^{-1}$. Used
parameters are $\eta=2$, $\lambda=3$, $x_{\mathrm{min}}=1$, $x_{\mathrm{max}}=1000$,
and $\sigma=1$.}
\label{fig:1}
\end{figure}

For $\lambda=3$ we get that $\beta=1$ and SDE \eref{eq:sde-ito}
gives signal exhibiting $1/f$ noise. Comparison of the numerically
obtained steady state PDF and the PSD with analytical expressions
for SDE \eref{eq:sde-ito} with $\eta=2$ and $\lambda=3$ is presented
in \fref{fig:1}. For the numerical solution we use the Euler-Maruyama
approximation, transforming the differential equations to difference
equations. We use a variable time step, decreasing with the increase
of $x$. As in \cite{Kaulakys2004,Kaulakys2006} we choose the time
step in such a way that the coefficient before noise becomes proportional
to the first power of $x$. Very similar numerical results one gets
also by using the Milstein approximation \cite{Kaulakys2009}. We
see a good agreement of the numerical results with analytical expressions.
A numerical solution of the equations confirms the presence of the
frequency region for which the power spectral density has $1/f^{\beta}$
dependence. The $1/f$ interval in the PSD in \fref{fig:1} is
approximately between $f_{\mathrm{min}}\approx10^{0}$ and
$f_{\mathrm{max}}\approx10^{3}$ and
is much narrower than the width of the region $1\ll f\ll10^{6}$ predicted
by equation \eref{eq:approx-range}. The width of this region can be
increased by increasing the ratio between the minimum and the maximum
values of the stochastic variable $x$. In addition, the region in
the power spectral density with the power-law behavior depends on
the exponent $\eta$: if $\eta=1$ then this width is zero; the width
increases with increasing the difference $|\eta-1|$ \cite{Ruseckas2010}.

\section{Nonlinear SDE resulting from motion in inhomogeneous medium}

\label{sec:Deriv_mod}

\begin{figure}
\includegraphics[scale=0.75]{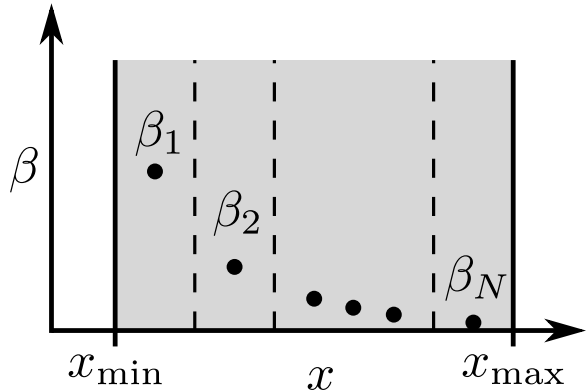}
\caption{The schematic representation of an inhomogeneous medium. The non-equilibrium
system is described as a large number $N$ of regions at local equilibrium,
each region having different inverse temperature $\beta_{i}$. When
the number of regions is sufficient large we can use approximation
\eref{eq:v_dist-formula}.}
\label{fig:superstat}
\end{figure}

The nonlinear SDE \eref{eq:sde-ito} generating signals with $1/f$
spectrum in a wide range of frequencies has been used so far to describe
socio-economical systems \cite{Gontis2010,Mathiesen2013}. The derivation
of the equations has been quite abstract and physical interpretation
of assumptions made in the derivation is not very clear. In this Section
we present a physical model where such equation can be relevant. We
expect that this derivation leads to a better understanding which
systems can be described using equation \eref{eq:sde-ito} or
equation \eref{eq:sde-stratonovich}.

We will consider the Brownian motion of a small macroscopic particle
in an inhomogeneous medium. We assume that this medium has reached
local thermodynamical equilibrium but not the global one and the temperature
can be considered as a function of coordinate. Schematically such
a medium is shown in \fref{fig:superstat}. Brownian motion of
small macroscopic particles in a liquid or a gas results from unbalanced
bombardments due to surrounding molecules. Usually the Brownian motion
is described by a Langevin equation that includes the influence of
the ``bath'' of surrounding molecules as a time-dependent stochastic
force that is commonly assumed to be a white Gaussian noise. This
assumption is is valid when the correlation time of fluctuations is
short, much shorter than the time scale of the macroscopic motion,
and the interactions with the bath are weak. The effect of large correlation
time of fluctuations will be considered in next Section. In the case
of strong collisions between the particle and the surrounding environment
the noise is non-Gaussian and we have so called L\'evy flights \cite{Weron2005}.
Nonlinear SDEs L\'evy noise and generating signals with $1/f$ spectrum
are proposed in \cite{Kazakevicius201X}. 

Langevin equations for one-dimensional motion of a Brownian particle
are \cite{Risken1989}
\begin{eqnarray}
\frac{\rmd}{\rmd t}v(t) = -\gamma v(t)+\frac{1}{m}F(x)
+\sqrt{\frac{2\gamma}{m}\frac{1}{\beta(x)}}\xi(t)\,,\label{eq:s-1}\\
\frac{\rmd}{\rmd t}x(t) = v(t)\label{eq:s-2}
\end{eqnarray}
Here $x$ is the coordinate and $v$ is the velocity of the Brownian
particle, $m$ is the mass of the particle, $\gamma$ is the relaxation
rate and $\xi(t)$ is the $\delta$-correlated white noise. In general,
equations similar to \eref{eq:s-1}, \eref{eq:s-2} can be used
to describe a variety of systems: noisy electronic circuits, laser
light intensity fluctuations \cite{Risken1989} and others. We assume
that there is temperature gradient in the medium, therefore the inverse
temperature $\beta(x)$ depends on coordinate $x$. In the case when
$\beta(x)\equiv\beta=\mathrm{const}$, equations \eref{eq:s-1}
and \eref{eq:s-2} describe Brownian motion in the medium with
constant temperature $T=k_{\mathrm{B}}^{-1}/\beta$, where $k_{\mathrm{B}}$
is Boltzmann constant. 

In high friction (also called overdamped) limit \cite{Gardiner2004}
the relaxation rate is large, $\gamma\gg|dv/dt|$. Performing adiabatic
elimination of the velocity as in \cite{Sancho1982}, we obtain the equation
\begin{equation}
\frac{\rmd}{\rmd t}x(t)=\frac{1}{\gamma m}F(x)
+\frac{1}{2\gamma m}\frac{\beta^{\prime}(x)}{\beta(x)^{2}}
+\sqrt{\frac{2}{\gamma m}\frac{1}{\beta(x)}}\xi(t)\,.
\label{eq:x_in_high_relax}
\end{equation}
Here $\beta^{\prime}(x)\equiv\rmd\beta(x)/\rmd x$. This SDE should be interpreted
according to the Stratonovich convention. Note, that the second term in the
right hand side of equation \eref{eq:x_in_high_relax} arises due to position
dependence of the stochastic force in equation \eref{eq:s-1} \cite{Sancho1982}.
The It\^o SDE corresponding to \eref{eq:x_in_high_relax} is
\begin{equation}
\rmd x = \frac{1}{\gamma m}F(x)\rmd t
+\sqrt{\frac{2}{\gamma m}\frac{1}{\beta(x)}}\rmd W_t\,.
\end{equation}
For calculating stationary distribution of position $x$ in high friction limit
we will use a Fokker-Planck equation instead of SDE~\eref{eq:x_in_high_relax}. The
Fokker-Planck equation corresponding to \eref{eq:x_in_high_relax} can be written as
\begin{equation}
\frac{\partial}{\partial t}P(x,t)=-\frac{1}{\gamma m}\frac{\partial}{\partial x}F(x)P(x,t)
+\frac{1}{\gamma m}\frac{\partial^{2}}{\partial x^{2}}\frac{P(x,t)}{\beta(x)}\,.
\end{equation}
By setting the time derivative to zero we obtain an analytical expression
for steady state distribution $P_{0}(x)$:
\begin{equation}
P_{0}(x)=C\beta(x)\exp\left(\int^{x}F(x')\beta(x')\rmd x'\right)\,.
\label{eq:SDist-Formula}
\end{equation}

Let us consider the situation when the dependence of the inverse temperature
on the coordinate is described by a power law,
\begin{equation}
\beta(x)=bx^{-\theta}\,.\label{eq:Intemp-form}
\end{equation}
Here $\theta$ is a power law exponent and $b$ is a constant. This
is quite reasonable assumption; for example, if $\theta=1$ equation \eref{eq:Intemp-form}
represents a case where we have steady state heat transfer due to temperature
difference $T_{2}-T_{1}$ between the beginning and the end of the system.
We assume that there are reflective boundaries at $x_{\mathrm{min}}$
and $x_{\mathrm{max}}$ and the motion is limited to values of $x$
between $x_{\mathrm{min}}$ and $x_{\mathrm{max}}$. When $\theta=1$
then the temperatures should obey the relation
$T_{2}/T_{1}=x_{\mathrm{max}}/x_{\mathrm{min}}$
and the coefficient $b$ is
$b=(x_{\mathrm{max}}-x_{\mathrm{min}})/k_{\mathrm{B}}(T_{2}-T_{1})$.
This case is presented in \fref{fig:superstat}.

The external force affecting the particle $F(x)$ we express via the
gradient of the potential $V(x)$:
\begin{equation}
F(x)=-\frac{\rmd}{\rmd x}V(x)\,.
\end{equation}
We choose the subharmonic potential proportional to the temperature:
\begin{equation}
V(x)=\left(\frac{\nu}{\theta}-1\right)\frac{1}{\beta(x)}\,.\label{Potential}
\end{equation}
For convenience we write the coefficient of proportionality as $\nu/\theta-1$.
As we will see in equation \eref{eq:S_dist_of_x}, the parameter $\nu$
gives the power law exponent in the steady state distribution of $x$.
Taking into account equation \eref{eq:Intemp-form} we see that the potential
has the power law form with the same exponent $\theta$. The expression
for the external force then is
\begin{equation}
F(x)=\frac{\theta}{b}x^{\theta-1}\left(1-\frac{\nu}{\theta}\right)\,.\label{eq:Force}
\end{equation}
Using inverse temperature \eref{eq:Intemp-form} and force \eref{eq:Force}
the equation \eref{eq:x_in_high_relax} for the particle coordinate
becomes
\begin{equation}
\frac{\rmd}{\rmd t}x(t)=\frac{\theta}{\gamma m b}\left(\frac{1}{2}
-\frac{\nu}{\theta}\right)x^{\theta-1}+x^{\frac{\theta}{2}}
\sqrt{\frac{2}{\gamma m\mathrm{b}}}\xi(t)\,.\label{eq:x_our_case}
\end{equation}
By using equations \eref{eq:SDist-Formula}, \eref{eq:Intemp-form}
and \eref{eq:Force} we obtain distribution of particles in high
friction limit 
\begin{equation}
P_{0}(x)=\frac{\nu-1}{x_{\mathrm{min}}^{1-\nu}
-x_{\mathrm{max}}^{1-\nu}}x^{-\nu}\,.\label{eq:S_dist_of_x}
\end{equation}
Calculating distribution of particles we assumed that there are
reflective boundaries at $x_{\mathrm{min}}$ and $x_{\mathrm{max}}$ and
the motion is limited to values of $x$ between $x_{\mathrm{min}}$
and $x_{\mathrm{max}}$. \Eref{eq:x_our_case} has the same
form as Stratonovich SDE \eref{eq:sde-stratonovich}.

\subsection{Numerical solution}

\begin{figure}
\includegraphics[width=0.33\textwidth]{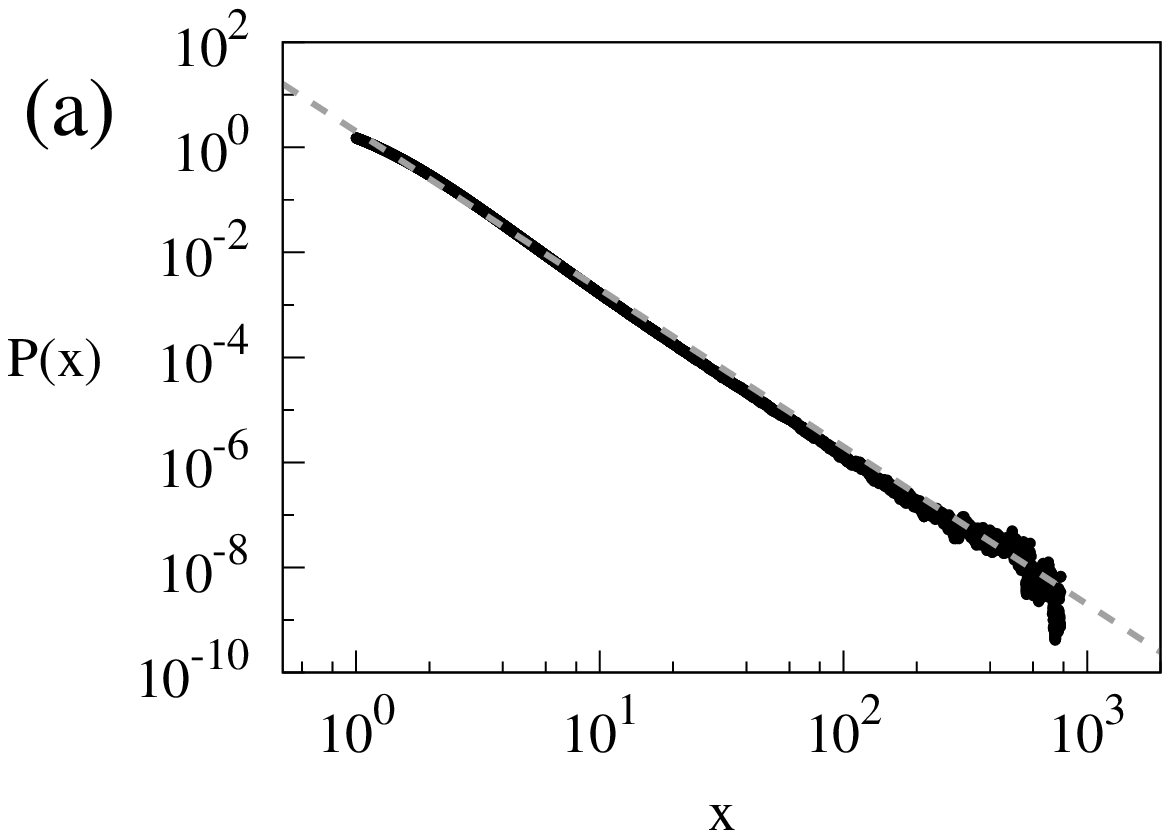}\includegraphics[width=0.33\textwidth]{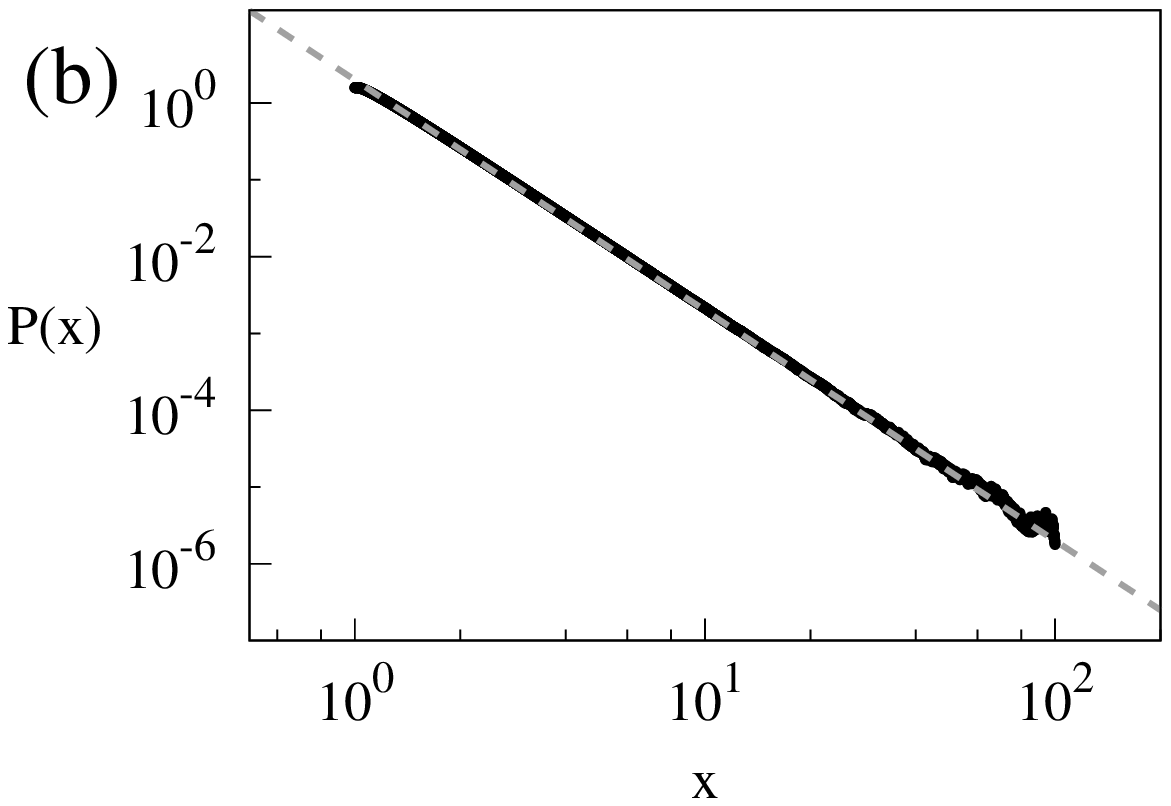}\includegraphics[width=0.33\textwidth]{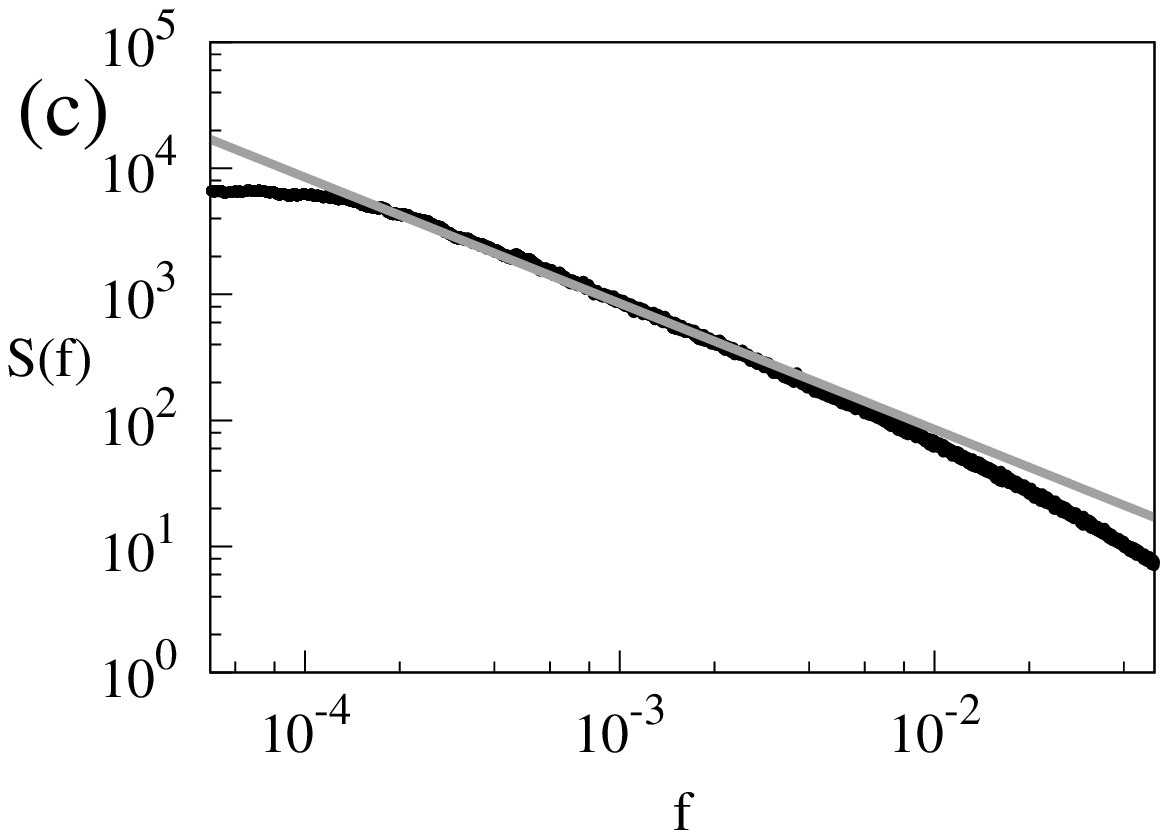}
\caption{The steady-state PDF of the signal generated by the Langevin equations
\eref{eq:dim-1}, \eref{eq:dim-2} with reflective boundaries at
$\tilde{x}_{\mathrm{min}}$ and $\tilde{x}_{\mathrm{max}}$ and the parameters:
(a) $\theta=1$, $\nu=3$, $\tilde{x}_{\mathrm{min}}=1$,
$\tilde{x}_{\mathrm{max}}=1000$, (b) $\theta=0$, $\nu=3$,
$\tilde{x}_{\mathrm{min}}=1$, $\tilde{x}_{\mathrm{max}}=100$. The dashed line
shows the power-law with the exponent $-3$. (c) The PSD of the signal
corresponding to the parameters in (b) case. The gray line shows the slope $f^{-1}$.}
\label{fig:langevin}
\end{figure}

To check the validity of the adiabatic elimination of the velocity, we solve the
Langevin equations \eref{eq:s-1}, \eref{eq:s-2} numerically. For numerical
solution it is convenient to introduce dimensionless time $\tilde{t}$, position
$\tilde{x}$ and velocity $\tilde{v}$. When the inverse temperature $\beta(x)$
and the force $F(x)$ are given by equations \eref{eq:Intemp-form} and
\eref{eq:Force}, we can use $\tilde{t}=\gamma t$ and
\begin{eqnarray}
\tilde{x} = (\gamma^{2}mb)^{\frac{1}{2-\theta}}x\,,\\
\tilde{v} = \gamma^{-1}(\gamma^{2}mb)^{\frac{1}{2-\theta}}v\,.
\end{eqnarray}
The equations \eref{eq:s-1}, \eref{eq:s-2} in the dimensionless variables have the form
\begin{eqnarray}
\frac{\rmd}{\rmd\tilde{t}}\tilde{v} = -\tilde{v}+(\theta-\nu)\tilde{x}^{\theta-1}
+\sqrt{2\tilde{x}^{\theta}}\xi(\tilde{t})\,,\label{eq:dim-1}\\
\frac{\rmd}{\rmd\tilde{t}}\tilde{x} = \tilde{v}\,.\label{eq:dim-2}
\end{eqnarray}
The reflective boundaries at $x_{\mathrm{min}}$ and $x_{\mathrm{max}}$ become
\begin{eqnarray}
\tilde{x}_{\mathrm{min}} = (\gamma^{2}mb)^{\frac{1}{2-\theta}} x_{\mathrm{min}}\,,\\
\tilde{x}_{\mathrm{max}} = (\gamma^{2}mb)^{\frac{1}{2-\theta}} x_{\mathrm{max}}\,.
\end{eqnarray}
The requirement of large friction $\gamma$, necessary for the adiabatic
elimination of the velocity, leads to the requirement
$\tilde{x}_{\mathrm{min}},\tilde{x}_{\mathrm{max}}\gg 1$ when $\theta < 2$.

Applying the Euler scheme with the step $h$ to \eref{eq:dim-1}, \eref{eq:dim-2} yields
the following equations:
\begin{eqnarray}
\tilde{v}_{k+1} = \tilde{v_k} -\tilde{v_k}h+(\theta-\nu)\tilde{x}_k^{\theta-1}h
+\sqrt{2\tilde{x}_k^{\theta}h}\xi_k\,,\\
\tilde{x}_{k+1} = \tilde{x}_k + \tilde{v}_k h\,.
\end{eqnarray}
Here $\xi_k$ are independent Gaussian random variables with zero mean and unit variance.

As an example we solve the Langevin equations \eref{eq:dim-1}, \eref{eq:dim-2}
with reflective boundaries at $\tilde{x}_{\mathrm{min}}$,
$\tilde{x}_{\mathrm{max}}$ and the parameters $\nu=3$, $\theta=1$ or $\theta=0$.
The value of the parameter $\theta=0$ means that the temperature is constant.
The steady state PDF of the particle position $P_0(\tilde{x})$ and the power
spectral density $S(\tilde{f})$ are presented in \fref{fig:langevin}. In
\fref{fig:langevin}a we see a good agreement with the distribution of particles
in high friction limit \eref{eq:S_dist_of_x}. This confirms the validity of the
adiabatic elimination of the velocity. \Fref{fig:langevin}c confirms the
presence of the frequency region with $1/f$ behavior of the power spectral
density, the $1/f$ interval in the PSD is approximately between
$\tilde{f}_{\mathrm{min}}\approx10^{-4}$ and
$\tilde{f}_{\mathrm{max}}\approx10^{-2}$. The width of this interval can be
increased by increasing the ratio
$\tilde{x}_{\mathrm{max}}/\tilde{x}_{\mathrm{min}}$. The width of $1/f^{\beta}$
region in the PSD also increases with increasing of $|\theta/2 -1|$.

Not only coordinate fluctuations yield power law PSD. The Langevin equation with
postion dependent temperature \eref{eq:Intemp-form} and external force
\eref{eq:Force} can also lead to the power law PSD of the absolute velocity
fluctuations.

\subsection{Superstatistics and velocity fluctuations}

In the motion of the particle there are two time scales: the scale
in which the particle is able to reach a local equilibrium and the
scale at which the fluctuating temperature changes. We assume that
temperature fluctuations are slow, that is, the time scale at which
observable change of temperature happens is much larger that relaxation
time of the particle $1/\gamma$. This assumption empowers us to derive
simpler equations for particle velocity and apply
the superstaticlical approach \cite{Beck2003,Tsallis2003,Abe2007,Hahn2010,Beck2011}
to obtain statistical properties of particle velocity fluctuations.
The superstatistical framework has been successfully applied to a
wide range of problems, such as interactions between hadrons from
cosmic rays \cite{Wilk2000}, fluid turbulence \cite{Beck2001,Beck2005,Beck2005a,Beck2007},
granular material \cite{Beck2006} and electronics \cite{Sattin2002}.
In the long-term the nonequilibrium system is described by the superposition
of different local dynamics at different time intervals. Superstatistics
is a description of the complex system under consideration by a superposition
of two statistics, one corresponding to ordinary statistical mechanics
(Langevin equation) and the other one corresponding to a slowly varying
parameter, in this case the inverse temperature $\beta(x)$. In the
superstatistical approach the distribution of the velocity of the
particle is 
\begin{equation}
P(v)=\sum_{i=1}^{N}f(\beta_{i})P(v|\beta_{i})\approx\int_{\beta_{\mathrm{min}}}^{\beta_{\mathrm{max}}}f(\beta)P(v|\beta)\rmd\beta\,,\label{eq:v_dist-formula}
\end{equation}
where $P(v|\beta)$ is a local velocity distribution for the particle
near the position $x$ when the temperature can be assumed to be constant
and equal to $\beta(x)=\beta$. For the Langevin equation in the form
of equation \eref{eq:s-1} the local velocity distribution is 
\begin{equation}
P(v|\beta)=\mathrm{\sqrt{\frac{m\beta}{2\pi}}}\exp\left(-\frac{1}{2}m\beta v^{2}\right)\,.\label{eq:Local_v_dist}
\end{equation}
The function $f(\beta)$ is the distribution of the inverse temperature.
The distribution of the inverse temperature $f(\beta)$ can be found
from the stationary distribution $P_{0}(x)$ of coordinate $x$ by
using the relation \cite{Straeten2012} 
\begin{equation}
f(\beta)=\left(\frac{\rmd\beta}{\rmd x}\right)^{-1}
P_{0}(x(\beta)).\label{eq:T_dist-formula}
\end{equation}
The parameters 
\begin{equation}
\beta_{\mathrm{max}}=bx_{\mathrm{min}}^{-\theta}\,,\qquad\beta_{\mathrm{min}}=\mathrm{b}x_{\mathrm{max}}^{-\theta}
\end{equation}
are maximal and minimal inverse temperatures. From equations \eref{eq:T_dist-formula},
\eref{eq:S_dist_of_x} and \eref{eq:Intemp-form} we obtain
the distribution of inverse temperature
\begin{equation}
f(\beta)=\frac{\nu-1}{\theta\left(\beta_{\mathrm{max}}^{\frac{\nu-1}{\theta}}-\beta_{\mathrm{min}}^{\frac{\nu-1}{\theta}}\right)}\beta^{\frac{\nu-1}{\theta}-1}\,.\label{eq:T_dist-result}
\end{equation}

Now we will describe fluctuations of particle velocity. For this purpose
we introduce the local average of absolute value of the velocity
\begin{equation}
\bar{v}\equiv\int\limits_{-\infty}^{\infty}|v|P(v|\beta)\rmd v\,.
\end{equation}
Keeping the assumption that local distribution of the velocity remains
Gaussian \eref{eq:Local_v_dist}, we have
\begin{equation}
\bar{v}=\sqrt{\frac{2}{\pi m\beta(x)}}=x^{\frac{\theta}{2}}
\sqrt{\frac{2}{\pi m\mathrm{b}}}\,.\label{eq:vid_mod_velosity}
\end{equation}
By using equation \eref{eq:vid_mod_velosity} and changing the variable
from $x$ to $\bar{v}$ in Stratonovich SDE \eref{eq:x_our_case}
\cite{Gardiner2004}, we obtain an SDE for the average velocity 
\begin{equation}
\frac{\rmd}{\rmd t}\bar{v}=\frac{1}{2}\sigma^{2}(\eta-\lambda)\bar{v}^{2\eta-1}
+\sigma\bar{v}^{\eta}\xi(t)\,.\label{eq:av_v_mod_result}
\end{equation}
Here we introduced the new parameters
\begin{equation}
\eta=2\left(1-\frac{1}{\theta}\right)\,,\qquad\lambda=\frac{2}{\theta}(\nu-1)+1
\end{equation}
and noise intensity
\begin{equation}
\sigma=\theta\sqrt{\frac{\pi}{4\gamma}}\left(\frac{2}{\pi mb}\right)^{\frac{1}{\theta}}\,.\label{eq:sk-parameter}
\end{equation}
The equation \eref{eq:av_v_mod_result} for the average absolute
velocity is identical to SDE \eref{eq:sde-stratonovich} generating
power-law distributed signals with $1/f^{\alpha}$ spectrum. The multiplicity
of noise $\eta$ in equation \eref{eq:av_v_mod_result} depends only
on power-law exponent $\theta$, whereas the exponent $\lambda$ of
the power law part of distribution of $\bar{v}$ depends on both parameters
$\theta$ and $\nu$. According to equation \eref{eq:beta} we have $1/f$
noise when $\lambda=3$ or $\nu=\theta+1$. One can expect that fast
velocity fluctuations do not influence the spectrum at small frequencies
and $1/f$ noise is visible not only in the spectrum of the local average
$\bar{v}$ but also in the spectrum of the absolute value of the velocity
$|v|$. Numerical calculation confirms this expectation.

From equations \eref{eq:v_dist-formula} and \eref{eq:T_dist-result}
we get
\begin{equation}
\fl P(v)=\frac{\lambda-1}{2\sqrt{\pi}\left(\beta_{\mathrm{max}}^{\frac{\lambda-1}{2}}
-\beta_{\mathrm{min}}^{\frac{\lambda-1}{2}}\right)}
\left(\frac{2}{m}\right)^{\frac{\lambda-1}{2}}|v|^{-\lambda}
\left[\Gamma\left(\frac{\lambda}{2},\frac{1}{2}mv^{2}\beta_{\mathrm{min}}\right)
-\Gamma\left(\frac{\lambda}{2},\frac{1}{2}mv^{2}\beta_{\mathrm{max}}\right)\right]
\label{eq:p-v}
\end{equation}
Here $\Gamma(a,z)=\int_{z}^{\infty}t^{a-1}\rme^{-t}\rmd t$ is the incomplete
gamma function. When $\frac{1}{2}mv^{2}\beta_{\mathrm{min}}\ll1$
and $\frac{1}{2}mv^{2}\beta_{\mathrm{max}}\gg1$ then from equation \eref{eq:p-v}
it follows that the distribution of velocities has a power law form
$P(v)\sim|v|^{-\lambda}$.

Thus we obtain $1/f$ noise in the fluctuations of the absolute value
of the velocity when the velocity distribution has a power-law part
$P(v)\sim v^{-3}$ and temperature distribution is flat, $f(\beta)=\mathrm{const}$.
When $\theta=1$ we have a simple system exhibiting $1/f$ fluctuations:
the system consists of a Brownian particle affected by a linear potential
$V(x)$ and moving in the medium where steady state heat transfer
is present due to the difference of temperatures at the ends of the
medium.

\begin{figure}
\includegraphics[width=0.45\textwidth]{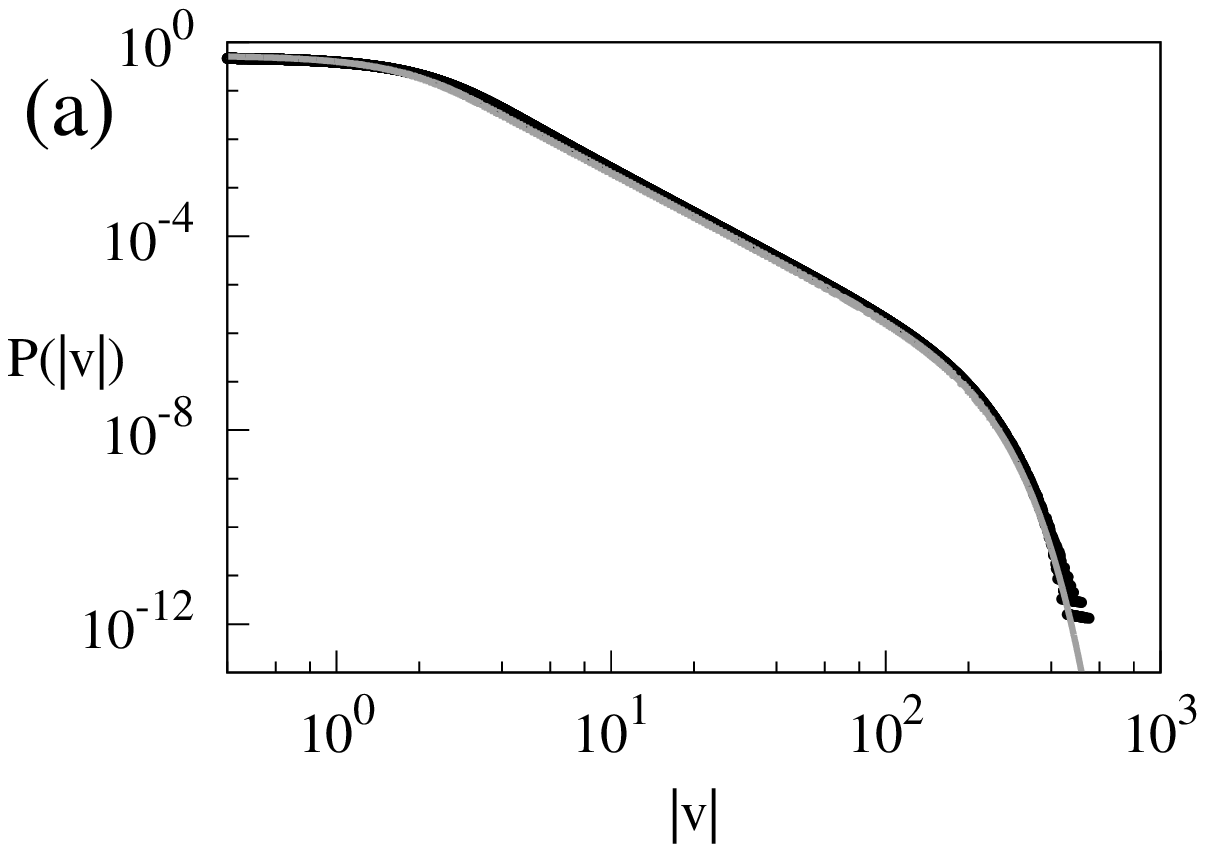}\includegraphics[width=0.453\textwidth]{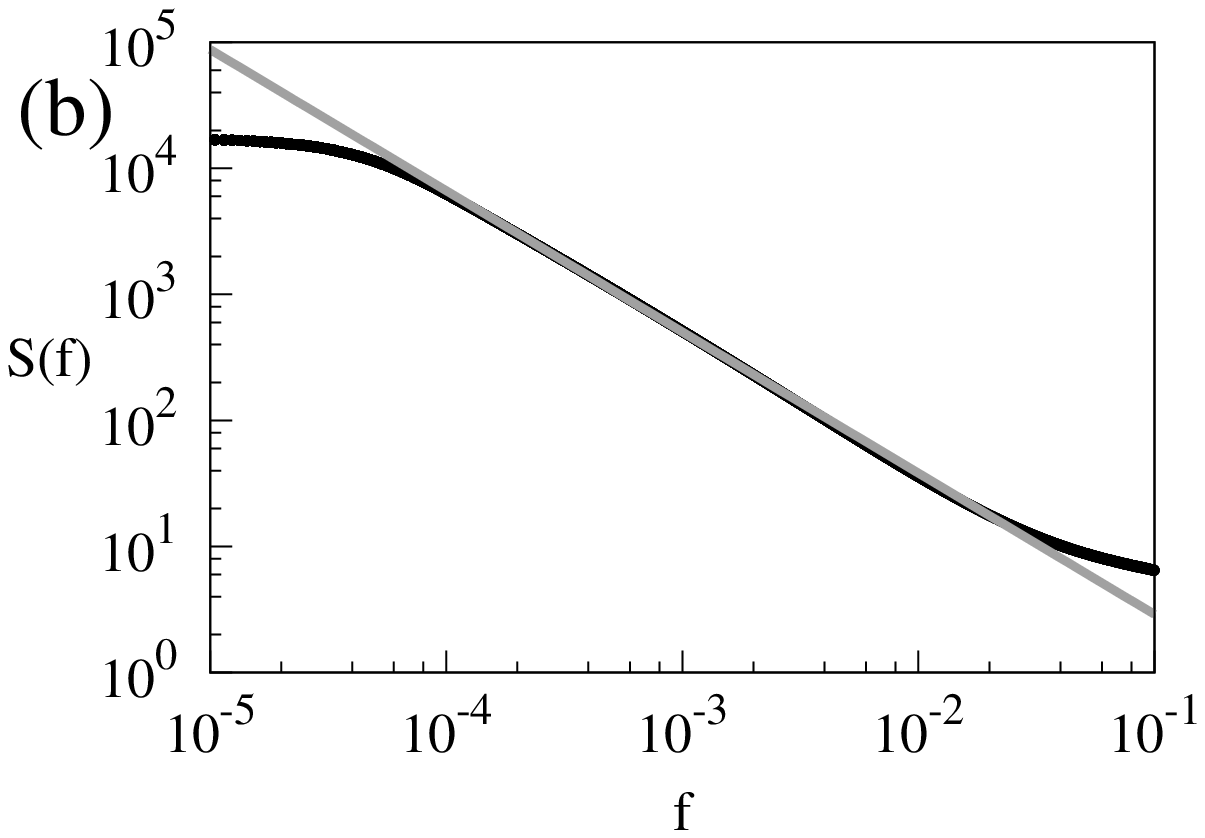}
\caption{(a) The steady-state PDF of the absolute velocity obtained by numerically
solving the Langevin equations \eref{eq:dim-1}, \eref{eq:dim-2} with reflective
boundaries at $\tilde{x}_{\mathrm{min}}$ and $\tilde{x}_{\mathrm{max}}$. The
gray line shows the PDF from equation \eref{eq:p-v} multiplied by 2. (b)
The PSD of the absolute velocity fluctuations. The gray line shows the slope
$f^{-1.1}$. The parameters used are: $\theta=1$, $\nu=2$,
$\tilde{x}_{\mathrm{min}}=1$, $\tilde{x}_{\mathrm{max}}=10^4$.}
\label{fig:langevin2}
\end{figure}

As an example we solve the Langevin equations \eref{eq:dim-1}, \eref{eq:dim-2}
with reflective boundaries at $\tilde{x}_{\mathrm{min}}$,
$\tilde{x}_{\mathrm{max}}$ and the parameters $\nu=2$, $\theta=1$. The steady
state PDF of the absolute velocity $P(|\tilde{v}|)$ and the power spectral
density $S(\tilde{f})$ of the absolute velocity fluctuations are presented in
\fref{fig:langevin2}. The steady state PDF $P(|v|)$ is twice as large as the PDF
given by equation \eref{eq:p-v} because both positive and negative velocities
with the same absolute value occur with equal probabilities. The PDF of the
dimensionless velocity $\tilde{v}$ can be obtained by setting $m=1$, $b=1$. In
\fref{fig:langevin2}a we see a good agreement with the analytical expression for
the steady state PDF. \Fref{fig:langevin2}b confirms the presence of the
frequency region with the power law behavior of the PSD. The power law region in
the PSD is approximately between $\tilde{f}_{\mathrm{min}}\approx10^{-4}$ and
$\tilde{f}_{\mathrm{max}}\approx10^{-2}$. However, there is a slight difference
between the power law exponent equal to $1$ predicted by analytical expressions
and the numerically obtained power law exponent $1.1$. This disagreement arises
because the analytical prediction of the power law exponent in the spectrum is
only approximate.

\section{Stochastic differential equations with colored noise}

\label{sec:colored} A system subjected to the colored noise is described
by the Langevin equation with a time-dependent stochastic force that
includes the influence of the ``bath'' of surrounding molecules:
\begin{equation}
\frac{\rmd x}{\rmd t}=f(x)+g(x)\varepsilon(t)\,.\label{gen_color_0}
\end{equation}
It is a well known result that if we approximate white noise by a
smooth, colored process, then at the limit as the correlation time
of the approximation tends to zero, the smoothed stochastic integral
converges to the Stratonovich stochastic integral \cite{Wong1965}.
We assume that the stochastic force in equation \eref{gen_color_0} is
a Gaussian noise having correlation time comparable with time scale
of the macroscopic motion and that $\varepsilon(t)$ is Markovian
process. In such a case the noise $\varepsilon$ satisfies Ornstein-Uhlenbeck
process with exponential correlation function \cite{Doob1944}. Thus
we describe the system perturbed by a colored noise as two-dimensional
Markovian flow:
\begin{eqnarray}
\frac{\rmd x}{\rmd t} = f(x)+g(x)\varepsilon\,,\label{gen_color_1}\\
\frac{\rmd\varepsilon}{\rmd t} = -\frac{1}{\tau}\varepsilon
+\frac{\sqrt{2D}}{\tau}\xi(t)\,.\label{gen_color_2}
\end{eqnarray}
Here $\xi(t)$ is the white noise, $\langle\xi(t)\xi(s)\rangle=\delta(t-s)$,
the parameter $\tau$ is the correlation time of the colored noise
and $D$ is the noise intensity. The autocorrelation function of the
colored noise is
\begin{equation}
\langle\varepsilon(t)\varepsilon(s)\rangle=\frac{D}{\tau}\exp\left(-\frac{|t-s|}{\tau}\right)\,.
\end{equation}

It is possible to write two dimensional Fokker-Planck equation for
equations \eref{gen_color_1}, \eref{gen_color_2} and obtain two dimensional
$P(x,\varepsilon)$ density as its solution. However, usually it is
enough to know the distribution $P(x)$ of the signal $x$, which
can be formally obtained by averaging $P(x,\varepsilon)$ over the
noise $\varepsilon$. It is problematic even get approximate analytical
solutions for $P(x,\epsilon)$ \cite{Hanggi1995}. More convenient
way is to get $P(x)$ from an approximate Fokker-Planck equation
just for one variable. Such equation can be obtained by using the
unified colored noise approximation \cite{Jung1987}, which is a form
of adiabatic approximation. To make this approximation we eliminate
the variable $\varepsilon$ and get the equation
\begin{equation}
\fl \frac{\rmd^{2}x}{\rmd t^{2}}=\frac{g^{\prime}(x)}{g(x)}
\left(\frac{\rmd x}{\rmd t}\right)^{2}-\left(\frac{1}{\tau}-f^{\prime}(x)
+f(x)\frac{g^{\prime}(x)}{g(x)}\right)\frac{\rmd x}{\rmd t}
+\frac{1}{\tau}f(x)+\frac{\sqrt{2D}}{\tau}g(x)\xi(t)\,.\label{Deriv_of g_c_1}
\end{equation}
We assume that the variable $x$ changes slowly and drop small terms
containing $\rmd^{2}x/\rmd t^{2}$ and $(\rmd x/\rmd t)^{2}$, obtaining the approximate
equation 
\begin{equation}
\frac{\rmd x}{\rmd t}=\frac{f(x)}{1-\tau\left(f^{\prime}(x)
-f(x)\frac{g^{\prime}(x)}{g(x)}\right)}+\frac{\sqrt{2D}g(x)}{1
-\tau\left(f^{\prime}(x)-f(x)\frac{g^{\prime}(x)}{g(x)}\right)}\xi(t)\,.\label{Unca_g_eq}
\end{equation}
This equation should be interpreted in the Stratonovich sense. In
more general case, when we cannot neglect inertia and drop the second
derivative $d^{2}x/dt^{2}$, the question whether the equation obtained
using adiabatic approximation should be interpreted in It\^o or Stratonovich
sense still remains an open question, so called It\^o-Stratonovich
problem \cite{Graham1982}. However, at least for specific systems
in white noise limit, it can be determined which interpretation is
correct. For example, it has been shown for a simplified model of the
preferential concentration of inertial particles in a turbulent velocity
field \cite{Sigurgeirsson2002}, that the equation obtained using
adiabatic elimination in white noise limit became the Stratonovich
equation \cite{Kupferman2004}. The Stratonovich interpretation should
be used if the correlation time of the noise is much larger than the
relaxation rate of the system. In an opposite case the equation should
be interpreted in It\^o sense. If relaxation rates are of the similar
magnitude as the correlation time, we get an equation with noise induced
drift that is different from Stratonovich drift.

The Fokker-Planck equation corresponding to the Stratonovich equation
$\rmd x/\rmd t=f_{\mathrm{c}}(x)+g_{\mathrm{c}}(x)\xi(t)$ is \cite{Gardiner2004}
\begin{equation}
\frac{\partial}{\partial t}P(x,t)=-\frac{\partial}{\partial x}f_{c}(x)P(x,t)
+\frac{1}{2}\frac{\partial}{\partial x}g_{c}(x)\frac{\partial}{\partial x}g_{c}(x)P(x,t)\,.\label{F_P_Ucna_gen}
\end{equation}
The applicability of equations \eref{Unca_g_eq} and \eref{F_P_Ucna_gen}
has limitation due to neglect of higher order derivatives in
equation \eref{Deriv_of g_c_1}.
These equations describe the dynamics correctly \cite{Jung1987} for
times $t$ obeying 
\begin{equation}
t\gg\frac{\tau}{1-\tau\left(f^{\prime}(x)-f(x)\frac{g^{\prime}(x)}{g(x)}\right)}
\end{equation}
and in the space regions obeying
\begin{equation}
\frac{\sqrt{2D\tau}g(x)}{1-\tau\left(f^{\prime}(x)-f(x)\frac{g^{\prime}(x)}{g(x)}\right)}\left|\frac{f^{\prime}(x)}{f(x)}\right|\ll1\,.\label{eq:validity-space}
\end{equation}

\section{Influence of colored noise on the stochastic differential equation
generating signals with $1/f$ spectrum}

\label{sec:col_on_model}If the nonlinear SDE generating signals with
$1/f$ spectrum is a result of a Brownian motion in an inhomogeneous
medium then the finite correlation time of the ``bath'' can become
important. In this section we use the results presented in \sref{sec:colored}
to investigate the influence of the colored noise. Instead of white noise
we add colored noise $\varepsilon(t)$ to the Stratonovich equation
\eref{eq:sde-stratonovich} obtaining the equations
\begin{eqnarray}
\frac{\rmd x}{\rmd t} = \frac{1}{2}\sigma^{2}(\eta-\lambda)x^{2\eta-1}
+\sigma x^{\eta}\varepsilon(t)\,,\label{eq:signal_w_color}\\
\frac{\rmd\varepsilon}{\rmd t} = -\frac{1}{\tau}\varepsilon
+\frac{1}{\tau}\xi(t)\,.\label{eq:40}
\end{eqnarray}
After unified colored noise approximation \eref{Unca_g_eq} we get 
\begin{equation}
\fl \frac{\rmd x}{\rmd t}=\frac{\frac{1}{2}\sigma^{2}(\eta
-\lambda)x^{2\eta-1}}{1-\frac{1}{2}\tau\sigma^{2}(\eta-1)(\eta-\lambda)x^{2(\eta-1)}}
+\frac{\sigma x^{\eta}}{1-\frac{1}{2}\tau\sigma^{2}(\eta-1)(\eta-\lambda)x^{2(\eta-1)}}\xi(t)\,.\label{eq:adiabat}
\end{equation}
If $\tau$ is large then equation \eref{eq:adiabat} has a simpler form
\begin{equation}
\frac{\rmd x}{\rmd t}=-\frac{x}{\tau(\eta-1)}+\frac{2x^{2-\eta}}{\tau\sigma(\eta-1)(\lambda-\eta)}\xi(t)\,.
\end{equation}
\Eref{eq:adiabat} should be interpreted in the Stratonovich
sense. Converting to It\^o interpretation \cite{Gardiner2004} we
have
\begin{equation}
\rmd x=\frac{1}{2}\frac{\sigma^{2}x^{2\eta-1}}{\gamma(x)}\left[\eta-\lambda
+\frac{2-\eta}{\gamma(x)}+\frac{2(\eta-1)}{\gamma(x)^{2}}\right]\rmd t
+\frac{\sigma x^{\eta}}{\gamma(x)}\rmd W_{t}\,,
\end{equation}
where
\begin{equation}
\gamma(x)\equiv1-\frac{1}{2}\tau\sigma^{2}(\eta-1)(\eta-\lambda)x^{2(\eta-1)}
\end{equation}
According to equation \eref{eq:validity-space}, approximation \eref{eq:adiabat}
is valid when
\begin{equation}
\frac{\sqrt{\tau}\sigma|2\eta-1|x^{\eta-1}}{1-\frac{1}{2}\tau\sigma^{2}(\eta-1)(\eta-\lambda)x^{2(\eta-1)}}\ll1
\end{equation}
Steady state PDF corresponding to equation \eref{eq:adiabat} is
\begin{equation}
\fl P_{0}(x)\sim x^{-\lambda}
\left(1-\frac{1}{2}\tau\sigma^{2}(\eta-1)(\eta-\lambda)x^{2(\eta-1)}\right)
\exp\left[-\frac{1}{4}\tau\sigma^{2}(\eta-\lambda)^{2}x^{2(\eta-1)}\right]
\label{eq:AdiabatPDF}
\end{equation}
We see that the colored noise introduces an exponential cut-off in
the steady state PDF $P_{0}(x)$ and naturally limits the range of
diffusion of the stochastic variable $x$. The exponential cut-off
is at large values of $x$ when $\eta>1$ and at small values of $x$
when $\eta<1$.

Comparing equation \eref{eq:adiabat} with \eref{eq:sde-stratonovich}
we see that the influence of the finite correlation time $\tau$ of
the noise can be neglected when
\begin{equation}
\frac{1}{2}\tau\sigma^{2}|\eta-1||\eta-\lambda|x^{2(\eta-1)}\ll1\,.\label{eq:cond-no-corr}
\end{equation}
Let us consider the case $\eta>1$. Then according to equation \eref{eq:cond-no-corr},
the influence of the finite correlation time $\tau$ of the noise
can be neglected when $x\ll x_{\tau}$, where
\begin{equation}
x_{\tau}\equiv\left[\frac{2}{\tau\sigma^{2}(\eta-1)|\eta-\lambda|}\right]^{\frac{1}{2(\eta-1)}}\label{eq:x-tau}
\end{equation}
If the diffusion is restricted to the region $x_{\mathrm{min}}<x<x_{\mathrm{max}}$
then the spectrum has a power-law part in the frequency range given
by \eref{eq:approx-range}. If $x_{\tau}>x_{\mathrm{max}}$
we expect no change in the power-law part of the spectrum. If $x_{\tau}<x_{\mathrm{max}}$
then, replacing the maximum value of $x$ by $x_{\tau}$ we get that
the replacement of the white noise by the colored noise leaves the
power-law part of the spectrum in the frequency range
\begin{equation}
\sigma^{2}x_{\mathrm{min}}^{2(\eta-1)}\ll2\pi f\ll\frac{2}{\tau(\eta-1)|\eta-\lambda|}\,.\label{eq:approx-range-1}
\end{equation}

If $\eta<1$ then the influence of the finite correlation time $\tau$
of the noise can be neglected when $x\gg x_{\tau}$, where $x_{\tau}$
is given by \eref{eq:x-tau}. When the diffusion is restricted
to the region $x_{\mathrm{min}}<x<x_{\mathrm{max}}$, we expect no
change in the power-law part of the spectrum when $x_{\tau}<x_{\mathrm{min}}$.
If $x_{\tau}>x_{\mathrm{min}}$ then replacing the minimum value of
$x$ by $x_{\tau}$ in equation \eref{eq:approx-range} we can estimate
that the power-law part of the spectrum should be in the frequency
range
\begin{equation}
\sigma^{2}x_{\mathrm{max}}^{-2(1-\eta)}\ll2\pi f\ll\frac{2}{\tau(1-\eta)|\eta-\lambda|}\,.
\end{equation}
Thus the introduction of the colored noise into equation \eref{eq:sde-stratonovich}
can narrow the range of frequencies where the PSD behaves as $1/f^{\beta}$
by decreasing the upper limiting frequency.

\subsection{Numerical solution}

\begin{figure}
\includegraphics[width=0.45\textwidth]{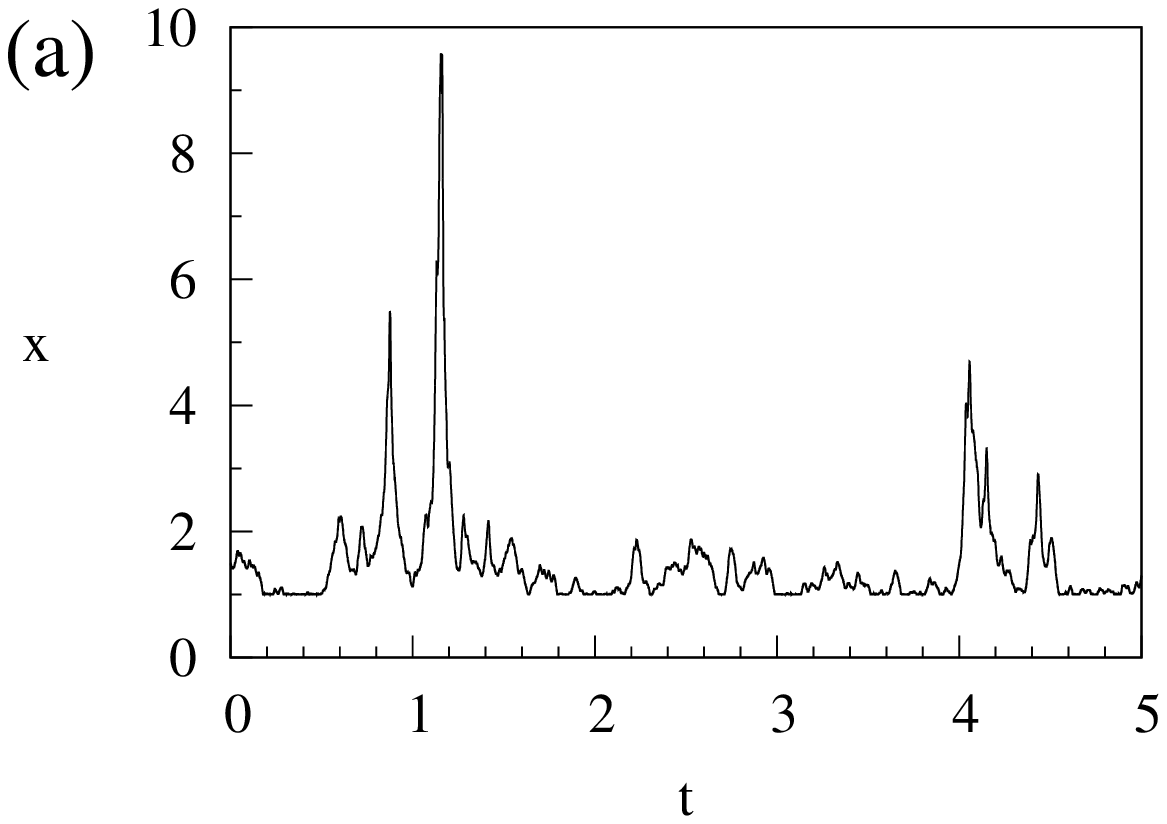}\includegraphics[width=0.45\textwidth]{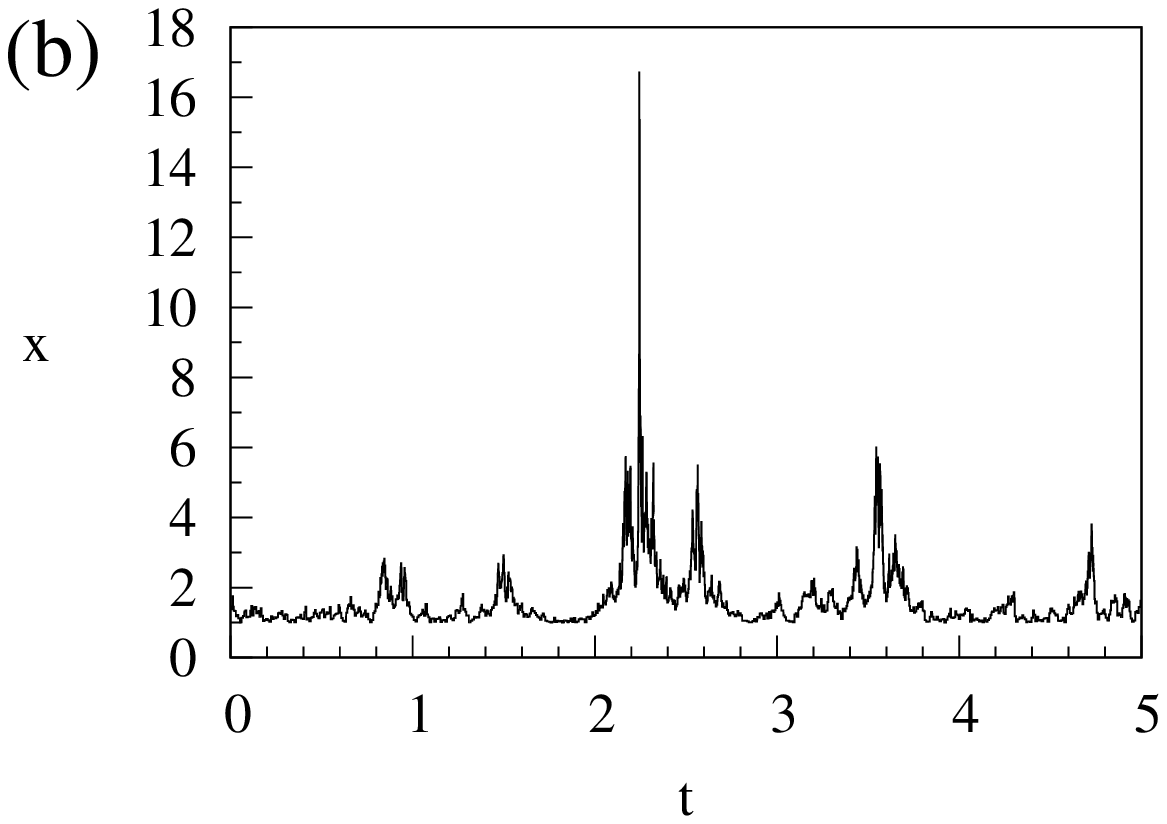}
\caption{(a) Typical signal generated by equations \eref{eq:signal_w_color},
\eref{eq:40} with $\tau=0.01$. (b) Typical signal generated by
equation \eref{eq:sde-stratonovich} corresponding to $\tau=0$ (white
noise). We used reflective boundaries at $x_{\mathrm{min}}=1$ and
$x_{\mathrm{max}}=1000$. Other parameters of equations are $\eta=2$,
$\lambda=3$, and $\sigma=1$.}
\label{fig:color-sig}
\end{figure}

\begin{figure}
\includegraphics[width=0.45\textwidth]{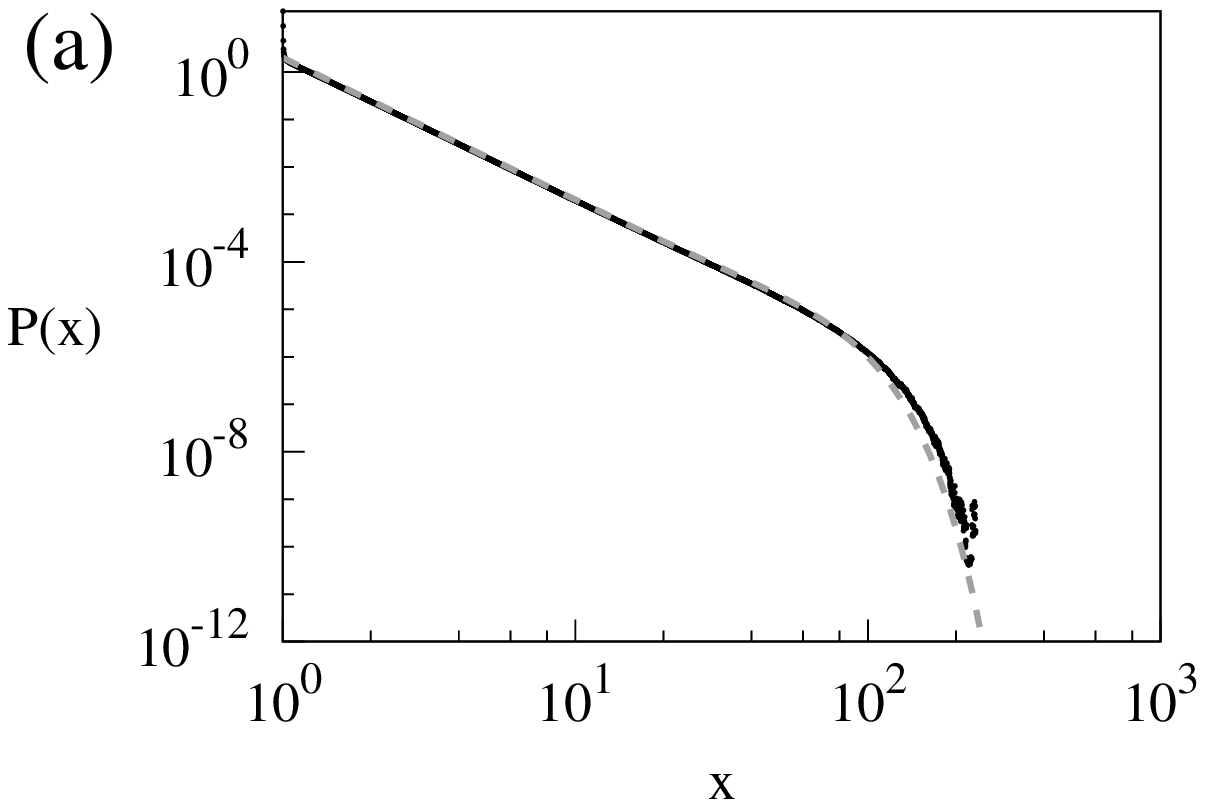}\includegraphics[width=0.45\textwidth]{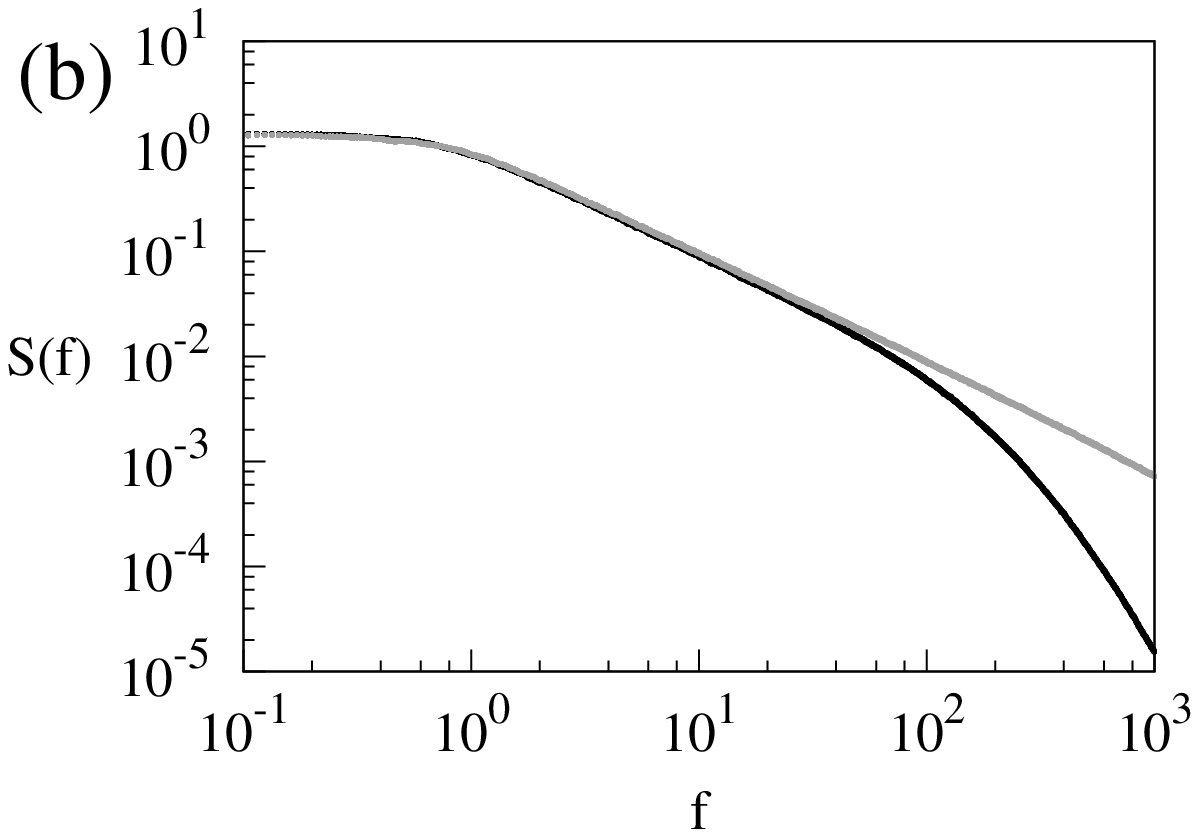}\\
\includegraphics[width=0.45\textwidth]{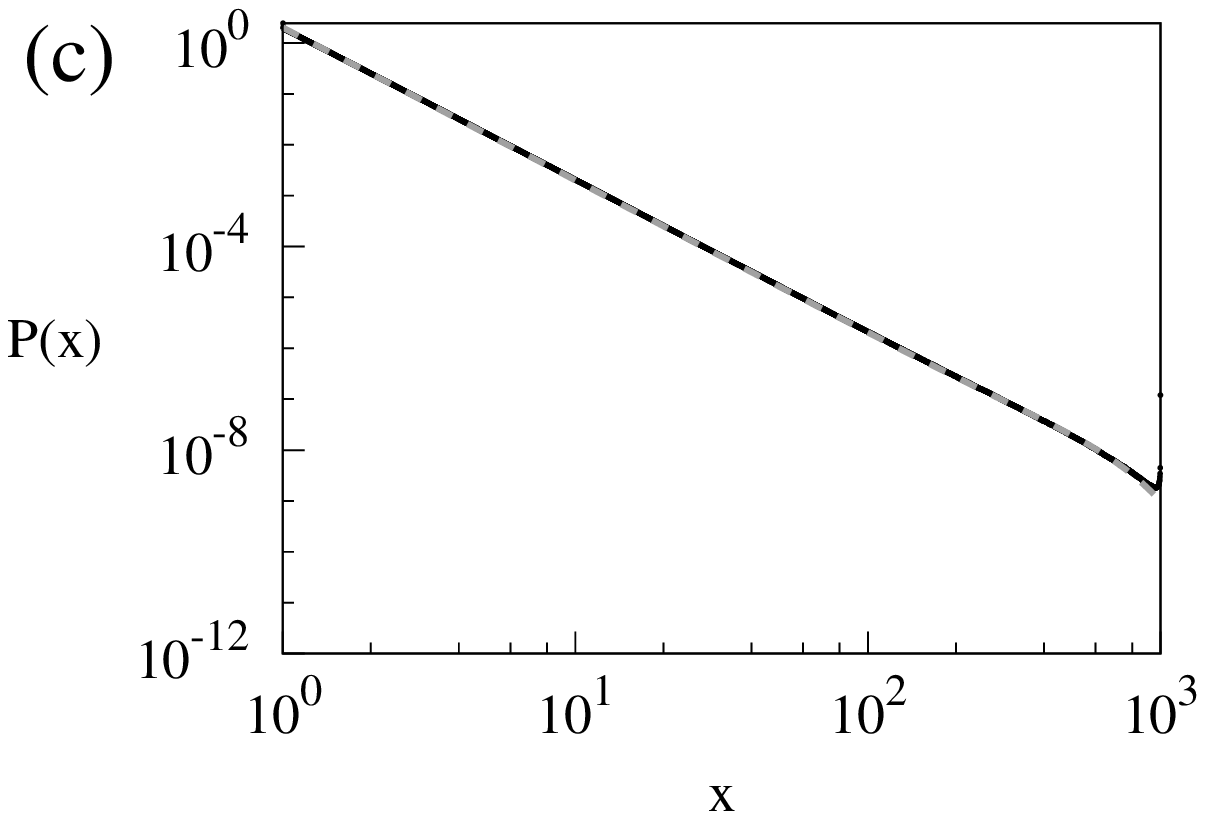}\includegraphics[width=0.45\textwidth]{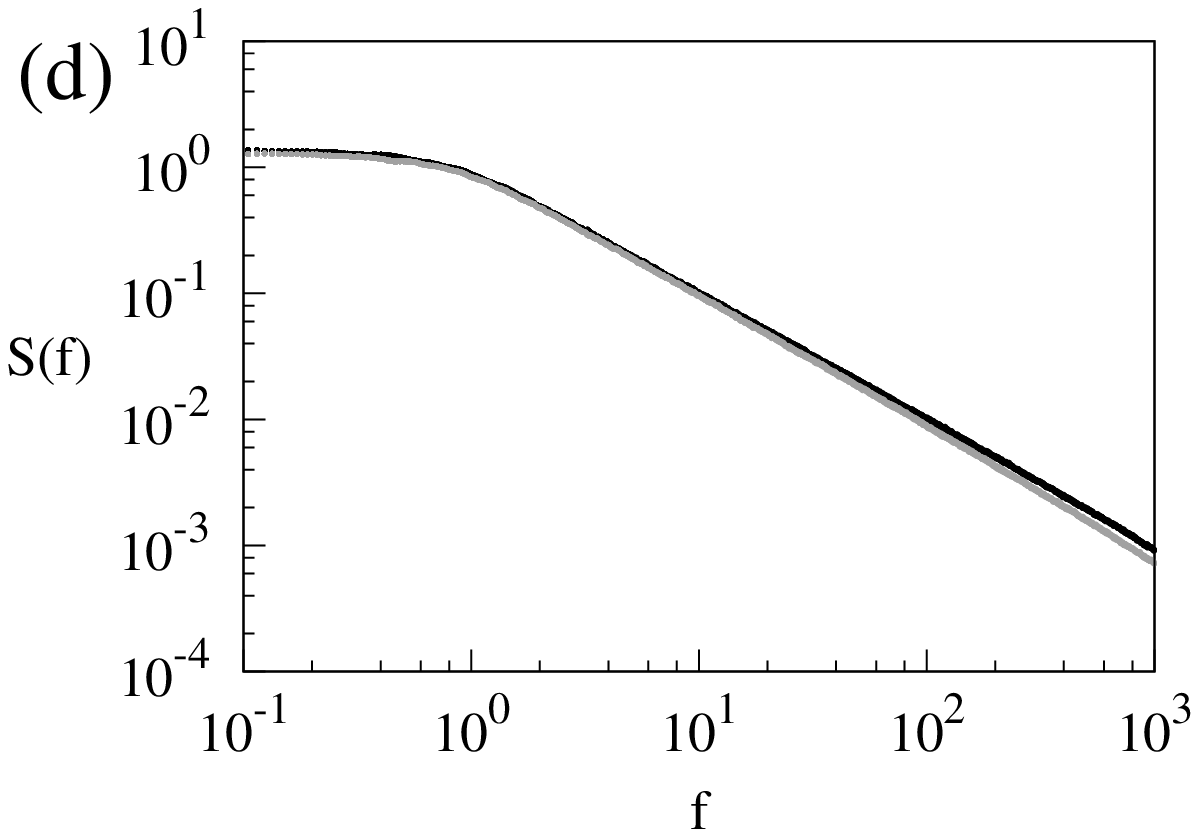}
\caption{(a,c) The steady-state PDF of the signal generated by
equations \eref{eq:Num_x_e_2})--\eref{eq:Num_t_e_2}
with reflective boundaries at $x_{\mathrm{min}}$ and $x_{\mathrm{max}}$.
The dashed line shows analytical approximation \eref{eq:AdiabatPDF}.
(b,d) The PSD of such a signal. The gray line shows the PSD of the
signal generated by equation \eref{eq:sde-stratonovich}. The correlation
time is (a,b) $\tau=10^{-3}$, (c,d) $\tau=10^{-5}$. Other parameters
are $\eta=2$, $\lambda=3$, $x_{\mathrm{min}}=1$, $x_{\mathrm{max}}=1000$,
and $\sigma=1$.}
\label{fig:color}
\end{figure}

To check the validity of the approximations we performed numerical
solution of equations \eref{eq:signal_w_color}, \eref{eq:40}. For
numerical solution we used the Euler scheme \cite{Fox1988,Milshtein1994}.
Applying the Euler scheme with the step $h$ to
\eref{eq:signal_w_color}, \eref{eq:40} yields the following equations:
\begin{eqnarray}
x_{k+1} = x_{k}+\left(\frac{1}{2}\sigma^{2}(\eta-\lambda)x_{k}^{2\eta-1}+\sigma x_{k}^{\eta}z_{k}\right)h\,,\label{eq:euler-1}\\
z_{k+1} = z_{k}-\frac{1}{\tau}z_{k}h+\frac{1}{\tau}\sqrt{h}\xi_{k}\,.
\end{eqnarray}
Here $\xi_{k}$ are independent Gaussian random variables with zero
mean and unit variance. For $\eta<1$ and for negative $\eta$ it
is convenient to use the Euler scheme with a constant time step $h$.
However, in the case of $\eta>1$ at large values of $x$ the coefficient
in equation \eref{eq:euler-1} becomes large and thus requires a very
small time step. A more effective solution is to use a variable time
step, decreasing with the increase of $x$, as has been done in Refs.~\cite{Kaulakys2004,Kaulakys2006}.
Variable time step $h_{k}=\kappa\tau/(\sigma x_{k}^{\eta})$ leads
to the equations
\begin{eqnarray}
x_{k+1} = x_{k}+\kappa\tau\left(\frac{1}{2}\sigma(\eta-\lambda)x_{k}^{\eta-1}+z_{k}\right)\,,\label{eq:var-1}\\
z_{k+1} = z_{k}-\frac{\kappa}{\sigma x_{k}^{\eta}}z_{k}+\sqrt{\frac{\kappa}{\sigma\tau x_{k}^{\eta}}}\xi_{k}\,,\\
t_{k+1} = t_{k}+\frac{\kappa\tau}{\sigma x_{k}^{\eta}}\,.\label{eq:var-3}
\end{eqnarray}
Here $\kappa\ll1$ is a small parameter.

As an example, we solve equations \eref{eq:signal_w_color}, \eref{eq:40}
with the parameters $\eta=2$, $\lambda=3$, $\sigma=1$ and reflective
boundaries at $x_{\mathrm{min}}=1$, $x_{\mathrm{max}}=1000$. From
equations \eref{eq:var-1}--\eref{eq:var-3} we get
\begin{eqnarray}
x_{k+1} = x_{k}+\kappa\tau\left(z_{k}-\frac{1}{2}x_{k}\right)\,,\label{eq:Num_x_e_2}\\
z_{k+1} = z_{k}-\frac{\kappa}{x_{k}^{2}}z_{k}+\frac{1}{x_{k}}\sqrt{\frac{\kappa}{\tau}}\xi_{k}\,,\label{eq:Num_z_e_2}\\
t_{k+1} = t_{k}+\frac{\kappa\tau}{x_{k}^{2}}\,.\label{eq:Num_t_e_2}
\end{eqnarray}
The generated signal is shown in \fref{fig:color-sig}. We see
that the finite correlation time $\tau$ of the noise leads to a smoother
signal compared to the equation with $\tau=0$. The steady state PDF
$P_{0}(x)$ and the power spectral density $S(f)$ for two different
values of $\tau$ are presented in \fref{fig:color}. From \fref{fig:color}a
we see that the unified colored noise approximation correctly predicts
the exponential cut-off in the steady state PDF at large values of
$x$, although the actual position of the cut-off slightly differs
from the cut-off predicted by equation \eref{eq:AdiabatPDF}. As \fref{fig:color}b
shows, the presence of the finite correlation time $\tau$ makes the
power-law part in the spectrum narrower. The upper limiting frequency
of the power-law region grows with decreasing of $\tau$, as is qualitatively
predicted by equation \eref{eq:approx-range-1}. The steady state PDF
and the PSD of the generated signal corresponding to much smaller
value of the correlation time $\tau$ are shown in \fref{fig:color}b,d.
For this value of $\tau$ the exponential cut-off due to finite correlation
time is larger than the upper boundary $x_{\mathrm{max}}$, thus we
see almost no differences from the case of uncorrelated noise, $\tau=0$.

\section{Discussion}

\label{sec:concl}

The nonlinear SDE~\eref{eq:sde-ito} generating signals with $1/f$
spectrum in a wide range of frequencies has been used so far to describe
socio-economical systems \cite{Gontis2010,Mathiesen2013}. The derivation
of the equations has been quite abstract and physical interpretation
of assumptions made in the derivation is not very clear. In this paper
we propose a physical model where such equations can be relevant.
This model is described by Stratonovich \eref{eq:sde-stratonovich}
instead of It\^o \eref{eq:sde-ito} SDE and provides insights which
physical systems can be described by such nonlinear SDEs.

We have shown that nonlinear SDEs generating power-law distributed
precesses with $1/f^{\beta}$ spectrum can result from diffusive particle
motion in inhomogeneous medium. The SDE~\eref{eq:av_v_mod_result}
for velocity fluctuations or SDE~\eref{eq:x_our_case} for particle
coordinate are simplified versions of Langevin equations \eref{eq:s-1},
\eref{eq:s-2} for one-dimensional motion of a Brownian particle.
We neglected viscosity dependence on temperature and inertia of particle.
In general, equations similar to these can be used to describe a variety
of systems: noisy electronic circuits, laser light intensity fluctuations
\cite{Risken1989} and others. We assumed that the inverse temperature
$\beta(x)$ depends on coordinate $x$ and this dependence is of a
power law form. Such a description is valid for a medium that has
reached local thermodynamical equilibrium but not the global one,
and the temperature can be considered as a function of coordinate.
The power law dependence of the inverse temperature $\beta(x)$ on the
stochastic variable $x$ can be caused by non-homogeneity of a bath.
This non-homogeneity can arise from a complex scale free structure
of the bath as is in the case of porous media \cite{Kulasiri2002}
or from the bath not being in an equilibrium.

In high friction limit, if the particle is affected by a subharmonic
potential proportional to the local temperature, the motion of the
particle can be described by the equation similar to equation \eref{eq:sde-stratonovich}.
For example, we can consider a Brownian particle affected by a linear
potential $V(x)$ and moving in the medium where steady state heat
transfer is present due to the difference of temperatures at the ends
of the medium. From the properties of equation \eref{eq:sde-stratonovich},
presented in \sref{sec:nonlinear-SDE}, it follows that the
spectrum of the fluctuations of the particle position $x(t)$ in such
a system can have a frequency region where the spectrum has a power-law
behavior. The width of this frequency region increases with the increase
of the length of the medium in which the particle moves.
\Eref{eq:sde-stratonovich}
can also describe the fluctuations of the local average of the absolute
value of the velocity, if temperature fluctuations are slow and the
superstatistical approach can be used. We obtain $1/f$ noise in the
fluctuations of the absolute value of the velocity when the velocity
distribution has a power-law part $P(v)\sim v^{-3}$ and temperature
distribution is flat, $f(\beta)=\mathrm{const}$.

The correlation of collisions between the Brownian particle and the
surrounding molecules can lead to the situation where the finite correlation
time becomes important, thus we have investigated the effect of colored
noise in our model. Using the unified colored noise approximation
we get that the finite correlation time leads to the additional restriction
of the diffusion. Existence of colored noise leads to an exponential
cut-off of the PDF of particle positions either from large values
when $\eta>1$, or from small values when $\eta<1$. Such a restriction
of the diffusion is a result of the multiplicative colored noise
in equation \eref{eq:signal_w_color}. Narrower power law part in the
PDF of the particle positions results in the narrower range of frequencies
where the spectrum has $1/f^{\beta}$ behavior. When $\eta>1$ , the
end of the power-law part of the spectrum at large frequencies is
inversely proportional to the correlation time $\tau$ of the noise.
However, for sufficiently small correlation time, when the restriction
of the diffusion due to colored noise is larger than the upper boundary
$x_{\mathrm{max}}$ of the medium, the effects of the colored noise
are negligible (see \fref{fig:color}) and the properties of
the signal do not differ significantly from the white noise case.

\section*{References}

\providecommand{\newblock}{}

\end{document}